# Widely Tunable Optical and Thermal Properties of Dirac Semimetal Cd$_3$As$_2$

*Hamid T. Chorsi[1], Shengying Yue[2], Prasad P. Iyer[1], Manik Goyal[3], Timo Schumann[3], Susanne Stemmer[3],*

*Bolin Liao[2], and Jon A. Schuller[1,\*]*

1. Department of Electrical and Computer Engineering, University of California, Santa Barbara, CA 93106, USA
2. Department of Mechanical Engineering, University of California, Santa Barbara, CA 93106, USA
3. Materials Department, University of California, Santa Barbara, CA 93106, USA

*Corresponding author (email: jonschuller@ece.ucsb.edu)

**Abstract**

In this paper we report a detailed analysis of the temperature-dependent optical properties of epitaxially grown cadmium arsenide (Cd$_3$As$_2$), a newly discovered three-dimensional Dirac semimetal. Dynamic Fermi level tuning—instigated from Pauli-blocking in the linear Dirac cone—and varying Drude response, generate large variations in the mid- and far-infrared optical properties. We demonstrate thermo-optic shifts larger than those of traditional III-V semiconductors, which we attribute to the obtained large thermal expansion coefficient as revealed by first-principles calculations. Electron scattering rate, plasma frequency edge, Fermi level shift, optical conductivity, and electron effective mass analysis of Cd$_3$As$_2$ thin-films are quantified and discussed in detail. Our *ab initio* density functional study and experimental analysis of epitaxially grown Cd$_3$As$_2$ promise applications for nanophotonic and nanoelectronic devices, such as reconfigurable metamaterials and metasurfaces, nanoscale thermal emitters, and on-chip directional antennas.

**Introduction**

Dirac semimetals present remarkable potential for high performance optical, electrical and thermal devices. The foremost example is graphene, a 2D Dirac semimetal, which has drawn tremendous attention since its discovery. Examples of graphene devices include electro-optic modulators [1, 2], field effect transistors (FETs) [3], spintronic devices [4], and tunable metasurfaces [5, 6]. Considerable interest in graphene applications and fundamentals has spurred a complementary interest in cadmium arsenide (Cd$_3$As$_2$)—a 3D Dirac semimetal. Cd$_3$As$_2$ possesses many of the same interesting optoelectronic properties as graphene with the additional benefit of possessing a strong light-matter interaction conferred by its three-dimensional nature [7, 8]. Ultrahigh mobility at room temperature, along with the large thermoelectric power factor [9], high carrier concentration [10], and resistance against oxidation [11], similarly make it a superb candidate for optoelectronic applications [7, 8, 12-14].

From a basic scientific perspective, Cd$_3$As$_2$ is an excellent playground for studying exotic and non-trivial topological phases of matter and intriguing condensed matter phenomena, such as linear quantum magnetoresistance [15] and chiral anomaly [16, 17]. Additionally, Cd$_3$As$_2$ has recently been exploited to realize a Weyl semimetal state [14-16]. To get a Weyl semimetal phase, either time reversal (TR) or inversion symmetry needs to be broken. Unlike graphene, the Dirac points in the Brillouin zone are protected by the point group ($C_4$ symmetry for $Cd_3As_2$) and cannot be gapped via spin-orbit coupling (SOC).

For future photonic, optoelectronic, and thermoelectric $Cd_3As_2$ devices, knowledge of the thermo-optic coefficient, the thermal expansion coefficient (TEC), and the carrier transport under different operating temperatures is crucial. To date, many tunable optical and thermal properties of $Cd_3As_2$ remain unexplored. Moreover, characterizing the dynamics of charge carriers subjected to electromagnetic perturbation is essential for the fundamental physics of optical excitations [18, 19].

In this paper, we use infrared (IR) spectroscopy to investigate the thermo-optic properties of $Cd_3As_2$ and demonstrate large optical tunability in the mid and far-IR regions. IR and THz spectroscopy are robust techniques for characterizing optical and electronic properties of Dirac and Weyl semimetals [20-24]. For example, IR spectroscopy of Dirac and Weyl semimetals has been used to study the excited transient excitonic instability [18] and chiral anomaly [17]. Through IR spectroscopy, we demonstrate large thermo-optic tuning of the $Cd_3As_2$ permittivity. Our modeling of these results supports the linear dispersion model for $Cd_3As_2$, providing estimates of the temperature-dependent electron scattering rate, Fermi energy, and plasma frequency of epitaxially-grown $Cd_3As_2$ films. Using *ab initio* density functional theory, we show that the large thermo-optic response derives from a large TEC. The results presented in this paper may excite further research into tunable optical and electronic response of Dirac and Weyl semimetals and may open up new applications in reconfigurable optoelectronic and nanoscale thermal devices.

**Film growth and characterization**

High quality $Cd_3As_2$ films were grown under ultra-high vacuum by molecular beam epitaxy (MBE) on (111)A-CdTe 4° miscut substrates (film thickness is 500 nm). Details of the growth can be found in our previous reports [25, 26]. $Cd_3As_2$ films display good surface morphology and high mobility as shown in Supplementary Figure S1. Infrared spectra were recorded using a Fourier Transform Infrared (FTIR, Vertex 70, Bruker) microscope with a 15X objective (N.A. 0.4) and liquid-nitrogen-cooled mercury–cadmium telluride (MCT) photodetector.

**Results and discussion**

The reflectivity spectra of an epitaxial (112)$Cd_3As_2$ thin-film grown on CdTe substrates are shown in **Figure 1** at various temperatures. At long wavelengths, a roll-off of the reflectivity due to the Drude response of free-carriers is observed. The roll-off becomes more pronounced with increasing temperature, indicating an increasing Drude response, i.e. an increase in plasma frequency with temperature. The effect on the infrared permittivity is indicated by a Kramers-Kronig (KK) consistent contribution to the permittivity describing this *intraband* response: $\varepsilon_{Drude} = -\frac{\omega_p^2}{\omega^2+i\Gamma\omega}$. To achieve good fits of the infrared permittivity an additional *interband* optical response is needed.

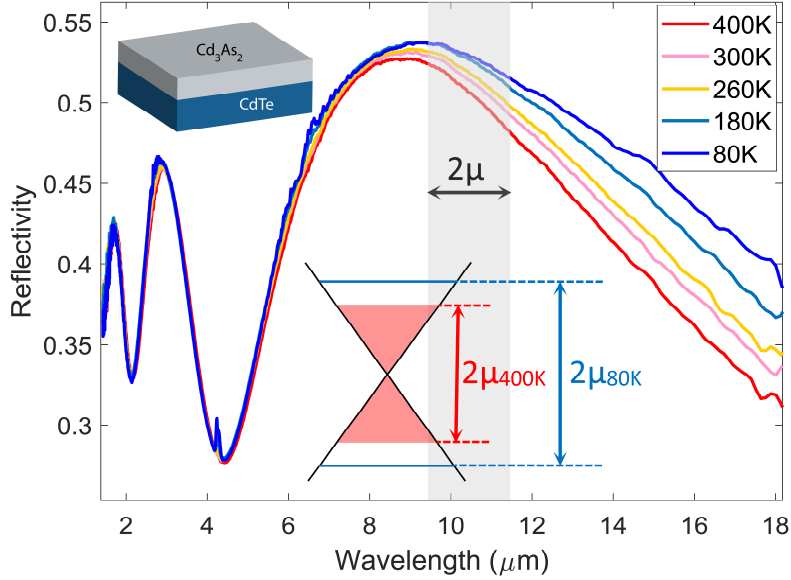

Figure 1. Reflectivity of Cd$_3$As$_2$ at different temperatures. Measurements are normalized to the reflectivity of gold.

Consider the body-centered tetragonal primitive unit cell of Cd$_3$As$_2$ [27], shown in **Figure 2(a)**. Applying first-principle density functional theory (DFT), we obtain crossed linear electronic bands near the Γ point as shown in Figure 2(a) [10, 27], giving rise to Dirac fermions. The interband absorption is described by an optical conductivity given by Equation 1 [28, 29].

$$\sigma_{inter}(\omega) = \frac{e^2 N_w}{12h}\frac{\omega}{v_f}\Theta[\hbar\omega - 2\mu] \quad (1)$$

where $\Theta$ is a scattering-broadened Heaviside function [22], $N_w$ is twice the number of spin-degenerate cones (taken to be $N_w$=4 here), $v_f$ is the Fermi velocity, $h$ is Plank's constant, $e$ is the electron charge, and $\mu$ is the electrochemical potential (Fermi level, E$_F$ is interchangeably used). This real-valued conductivity produces an imaginary component of the permittivity according to $Im\{\varepsilon_{inter}\} = i\omega\sigma_{inter}(\omega)$. In turn, this produces a real valued permittivity component derived from a Kramers-Kronig transformation: $Re\{\varepsilon_{inter}(\omega)\} = -\frac{1}{\pi}P\int_{-\infty}^{\infty}\frac{Im\{\varepsilon_{inter}(\omega')\}}{\omega'-\omega}d\omega'$. Taking the intra and interband terms together we fit measured infrared reflectivity using a permittivity function described in Equation 2:

$$\epsilon = \epsilon_r + Re\{\varepsilon_{inter}\} + iIm\{\varepsilon_{inter}\} - \frac{\omega_p^2}{\omega^2+i\Gamma\omega} \quad (2)$$

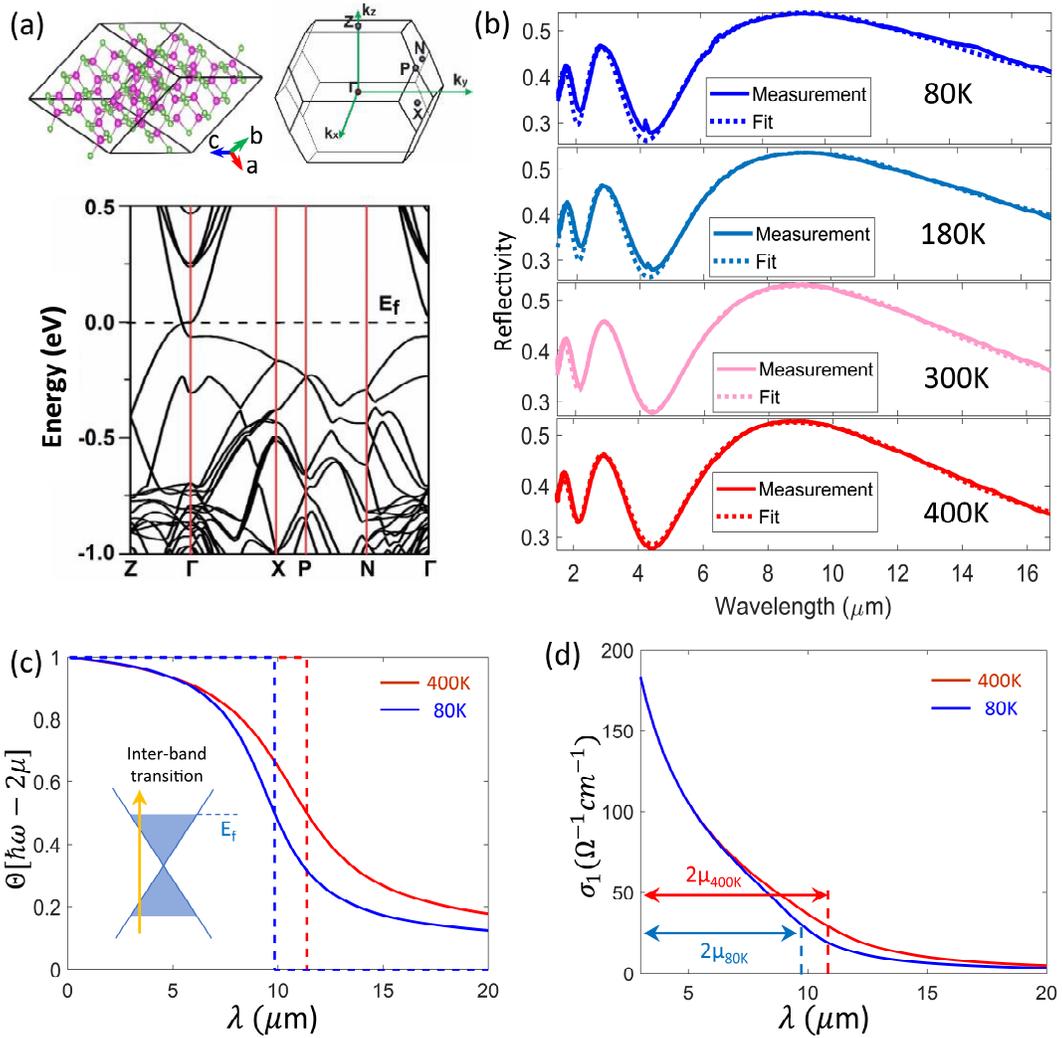

Figure 2. (a) The electronic band structure from DFT calculation with Perdew-Burke-Ernzerhof (PBE) exchange-correlation functional with SOC. The band structure for $Cd_3As_2$ with $I4_1/acd$ space group and 80 atoms per unit cell. (b) Fits to the experimental reflectivity curves in Figure 1 via KK consistent data. (c) Heaviside step function and broadening effect. Inset shows the Pauli blocking edge and the interband transition (orange arrow). (d) Optical conductivity as a function of wavelength at 80K and 400K.

Inputting this complex permittivity into transfer-matrix calculations, we fit the associated spectral reflectance curves with free-parameters: $\{v_f, \mu, \omega_p, \Gamma\}$. The temperature-dependent broadened Heaviside functions and associated optical conductivities are shown in **Figure 2(c) and 2(d)** respectively. The onset of interband absorption (>2μ) shifts to longer wavelengths (lower energy) as temperature increases, consistent with previous theoretical calculations of the temperature-dependent chemical potential [23,34]. A summary of temperature-dependent fit parameters is listed in Table 1 and summarized in **Figure 3**.

Table 1. Fitted parameters for the acquired temperature-dependent reflectivity data.

| Parameter | Value at 80K | Value at 180K | Value at 300K | Value at 400K |
|---|---|---|---|---|
| Thickness | 500 nm | 500 nm | 500 nm | 500 nm |
| $v_f$ | $1.6\times10^5$ m/s | $1.7\times10^5$ m/s | $2.3\times10^5$ m/s | $2.6\times10^5$ m/s |
| $\omega_p$ | 19.72 THz | 26.76 THz | 48.35 THz | 57.65 THz |
| $\Gamma$ | 3.92 THz (130.8 cm$^{-1}$) | 4.11 THz (137.2 cm$^{-1}$) | 4.40 THz (146.9 cm$^{-1}$) | 4.48 THz (149.5 cm$^{-1}$) |
| $2\mu$ | 126 meV | 124.5 meV | 115.6 meV | 109 meV |

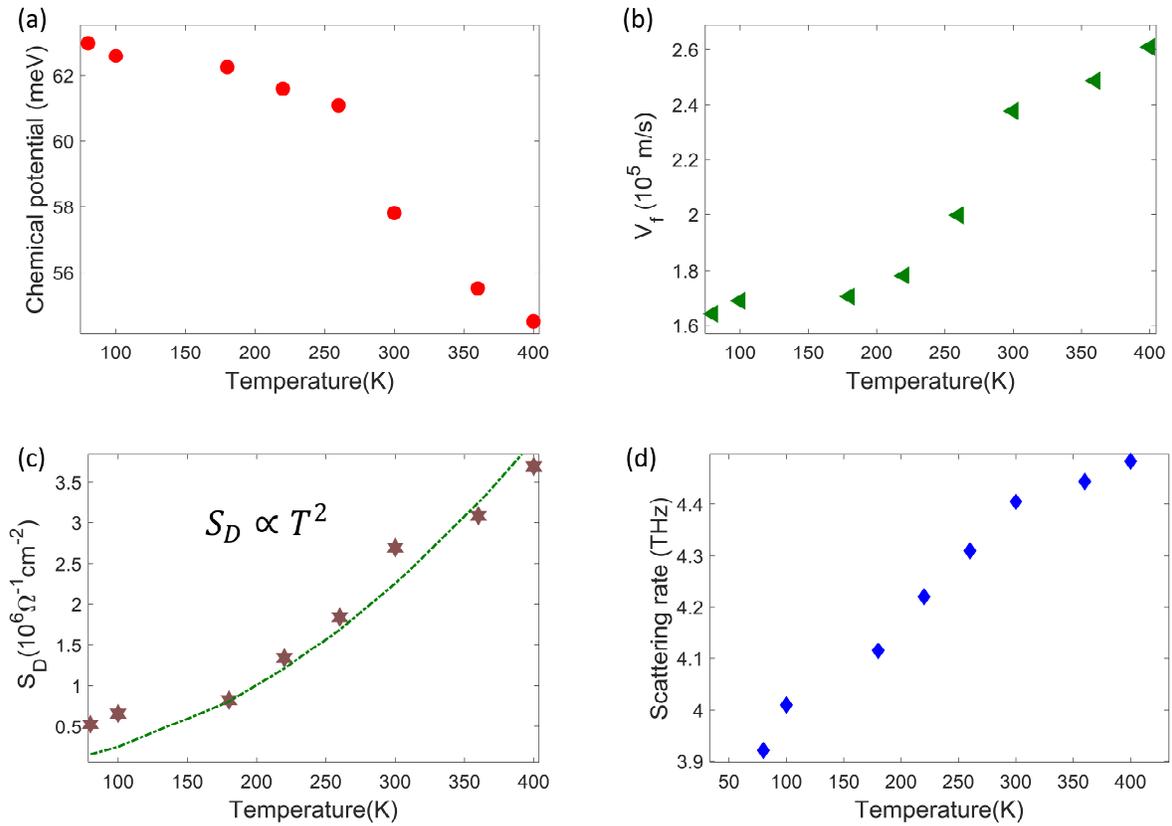

Figure 3. (a) Temperature dependent change of the Fermi level. (b) Fermi velocity variations from 80 K to 400 K. (c). Drude weight variations as a function of temperature. Dashed green line is the $\propto T^2$ proportionality. (d) Electron scattering rate changes with temperature. Markers indicate experimentally obtained data points.

The shifts in chemical potential shown in **Figure 3(a)** are complemented by an increase in Fermi velocity with increasing temperature **Figure 3(b)**. Our derived values for $v_f$ (1.6 x $10^5$ – 2.6 x $10^5$ m/s) are less than those determined from previous ARPES and STM measurements [13, 30] (7.6 x $10^5$ – 1.5 x $10^6$ m/s) but consistent with previous results derived from optical spectroscopy, and the trend with temperature follows expected behavior given the temperature-dependent chemical potential [20].

The temperature-dependent Cd$_3$As$_2$ Drude response is plotted in **Figure 3(c)** in terms of a Drude weight [35]:

$$S_D = 2\pi^3 c^2 \epsilon_0 \omega_p \qquad (3)$$

where $c$ is the speed of light in vacuum, $\epsilon_0$ is the permittivity of free space, and $\omega_p$ is the plasma frequency. The Drude weight exhibits a quadratic temperature dependence in good agreement with previous studies [31, 32]. A small increase in scattering rate with temperature is also observed **Figure 3(d)** and is qualitatively consistent with expected behavior in other Drude systems [33, 34].

The complex temperature-dependent permittivities inferred from these fits are plotted in **Figure 4(a)**. The average thermo-optic coefficient (TOC) from 80 to 400 K, (i.e., $\frac{\partial \epsilon}{\partial T}$), is 102 $\times 10^{-4}/K$ at a wavelength of 14 μm which is larger than the TOC of traditional III-V semiconductors [35]. A comparison between the TOC of Cd$_3$As$_2$ with a few well-known semiconductors is provided in the supplementary information (Supplementary Table S1). These results are compared to temperature-dependent optical permittivities calculated with the independent-particle approach (IPA) based density functional perturbation theory (DFPT), where the temperature dependence is captured by a quasi-harmonic approximation (QHA) calculation of the temperature-dependent lattice parameters. As mentioned above, the linear band structure of Cd$_3$As$_2$ can be obtained using the large unit cell model (80 atoms per unit cell), however, such a large unit cell is computationally intensive for the QHA calculations. We instead use a small 10-atom unit cell [10, 27] that shows good agreement with the more rigorous calculations (**Figure 4 (f)**), thanks to the fact that the long-wavelength IR response is not sensitive to the short-range arrangement of the Cd vacancies in the crystal structure. For each set of temperature-dependent lattice parameters we calculate and plot the average of the diagonal elements of the dielectric tensor. The real and imaginary parts are plotted in **Figure 4(b)**. The DFT results agree well with the attained experimental data.

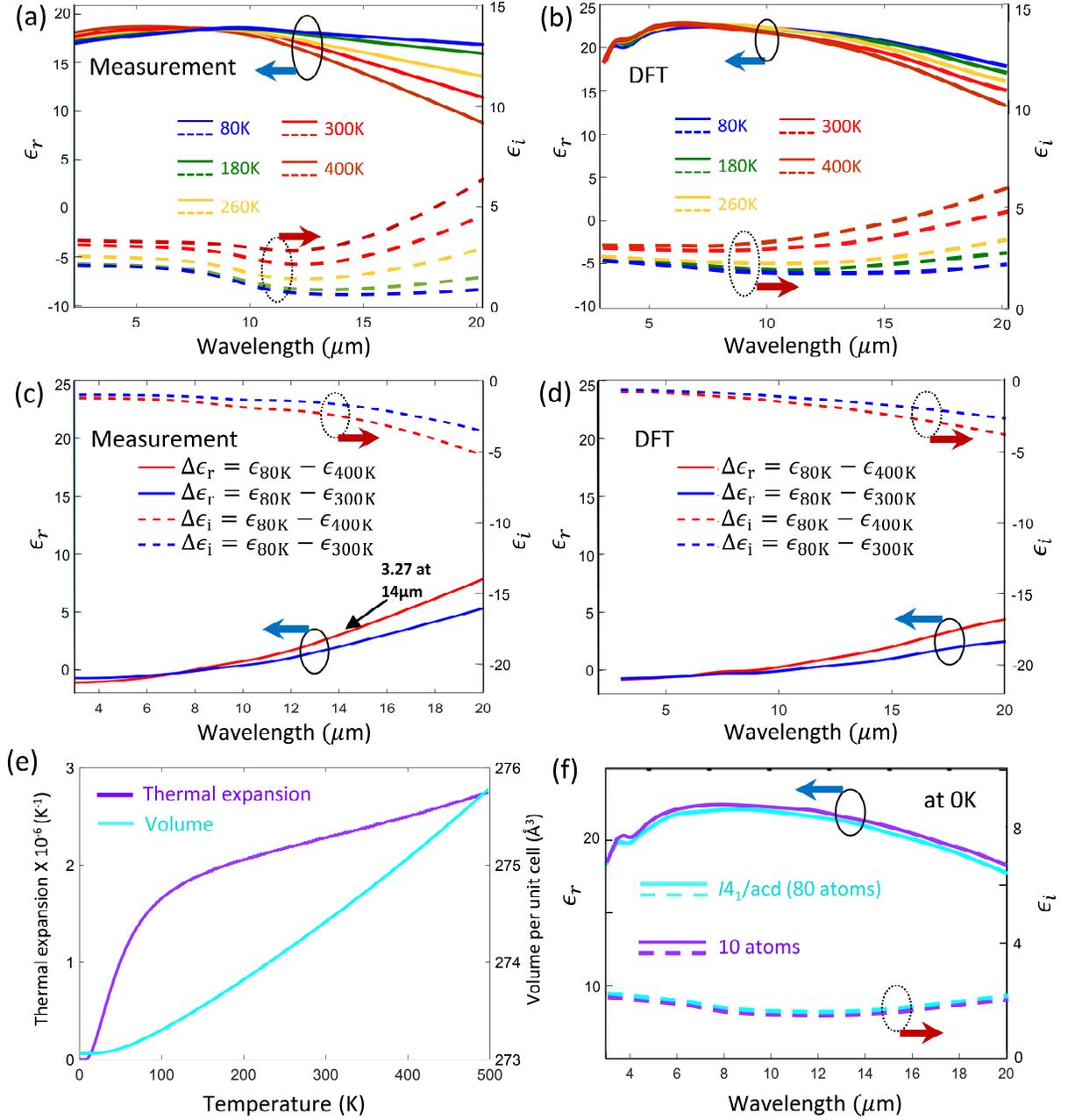

Figure 4. The real and imaginary parts of the permittivity with the Drude absorption included, (a) obtained via fitting experimental data, (b) DFT results. Temperature-dependent change of the real and imaginary parts of the permittivity in the IR region, (c) obtained via fitting experimental data, (d) DFT results. Real part: solid line; imaginary part: dashed line. Calculated TOC is 102 at 14μm. (e) TEC and temperature-dependent unit cell expansion of Cd$_3$As$_2$ obtained via DFT. (f) Wavelength-dependent permittivity for small (10 atoms per unit cell) versus large (80 atoms per unit cell) unit cells.

**Figure 4(e)** shows the large thermal expansion coefficient obtained from the QHA calculations. The large TOC in Cd$_3$As$_2$ is likely an effect of the large TEC, which is an order of magnitude higher than conventional III-V semiconductors. Thermal expansion of Cd$_3$As$_2$ results from weaker atomic binding between Cd and As at higher temperatures, which may induce greater polarizability of the lattice.

DFT-based results were performed for Cd$_3$As$_2$ as an intrinsic semimetal; however, the experimentally measured chemical potential was approximately 60 meV above the Dirac node, indicating an unintentionally n-type doped Cd$_3$As$_2$. This factor may be responsible for subtle differences between the physical measurements and DFT calculations.

The observed large changes in optical permittivity may be exploited to engineer reconfigurable infrared photonic elements such as optical antennas and metasurfaces. As a case study, using the results presented here we demonstrate simulations of reconfigurable absorption in a Cd$_3$As$_2$ metasurface array results can be found in the Supplementary information Figure S3.

**Conclusion**

In conclusion, we have studied the temperature-dependent optical properties of epitaxially grown three-dimensional Dirac semimetal Cd$_3$As$_2$ and demonstrated large IR thermo-optic shift originating from tuning of the Fermi distributions. The large thermo-optic shifts agree well with DFT calculations and likely arise from the materials' large thermal expansion coefficient. The results demonstrated here suggest the potential for creating reconfigurable electronic and optical devices based on Dirac and Weyl materials.

**Computational details:** Electronic properties calculations: for the small unit cell results we adopted a unit cell with 4 As atoms and 6 Cd atoms. The primitive cell had a tetragonal structure, in which the As atoms occupy corner and face-center positions and the 6 Cd atoms form a cube with two vacancies located diagonally on one surface, as shown in the inset of Supplementary Figure S4. The electronic band structures are calculated using Vienna ab-initio simulation package (VASP) [36, 37] with the Perdew-Burke-Ernzerhof (PBE) generalized gradient approximation (GGA) [38] exchange-correlation functional. The crystal structures were relaxed and the Hellmann–Feynman force tolerance was set to $10^{-6} eV Å^{-1}$. The kinetic energy cutoff for the wave functions was set at 680 eV (50 Ry), the energy convergence threshold was set as $10^{-8}$ eV. The Monkhorst-Pack k-meshes [39] of $4 \times 4 \times 4$ was used to sample the Brillouin zone. The convergence for the cutoff energy of the planewave basis and the k-grid density was checked. All the DFT calculations included the spin-orbit coupling (SOC) effects. Quasi-harmonic approximation calculation based on DFT was conducted using the PHONOPY code [40] to calculate the phonon and the thermal expansion properties. We used the finite small-displacement method with the atomic displacement amount 0.03Å. The supercells for these calculations are 3 × 3 × 3 (270 atoms).

*The static dielectric properties based on DFPT*: The average of the frequency-dependent dielectric function tensor was calculated within the independent-particle approximation (IPA) [41]. The calculations for the static dielectric properties were done using the density functional perturbation theory (DFPT). The temperature-dependence originated from the varying lattice parameters at different temperature from the QHA calculations. For the dielectric function calculations, we adopted a fine k-mesh 10 × 10 × 10, and denser meshes were used around the Dirac nodes for added resolution for the long wavelength region.

**Acknowledgment**

The authors thank David Kealhofer and Omor Shoron for very useful discussions. This work was supported by the Air Force Office of Scientific Research (Grant No. FA9550-16-1-0393). M. G., T.S. and S.S. acknowledge funding by the Vannevar Bush Faculty Fellowship program of the U.S. Department of Defense (grant no. N00014-16-1-2814). The DFT simulation was supported by the Department of Energy, Office of Basic Energy Sciences through the Early Career Research Program under the award number DE-SC0019244. We also acknowledge support from the Centre for Scientific Computing from the CNSI and NSF Grant No. CNS-0960316.

Received: ((will be filled in by the editorial staff))
Revised: ((will be filled in by the editorial staff))
Published online: ((will be filled in by the editorial staff))

**Conflict of Interest**

The authors declare no conflict of interest.

**References**

1. Liu, M.; Yin, X.; Ulin-Avila, E.; Geng, B.; Zentgraf, T.; Ju, L.; Wang, F.; Zhang, X. *Nature* **2011,** 474, 64.
2. Bao, Q.; Loh, K. P. *ACS Nano* **2012,** 6, (5), 3677-3694.
3. Schwierz, F. *Nature Nanotechnology* **2010,** 5, 487.
4. Han, W.; Kawakami, R. K.; Gmitra, M.; Fabian, J. *Nature Nanotechnology* **2014,** 9, 794.
5. Yao, Y.; Shankar, R.; Kats, M. A.; Song, Y.; Kong, J.; Loncar, M.; Capasso, F. *Nano Letters* **2014,** 14, (11), 6526-6532.
6. Chorsi, H. T.; Gedney, S. D. *IEEE Photonics Technology Letters* **2017,** 29, (2), 228-230.

7.	Wang, Q.; Li, C.-Z.; Ge, S.; Li, J.-G.; Lu, W.; Lai, J.; Liu, X.; Ma, J.; Yu, D.-P.; Liao, Z.-M.; Sun, D. *Nano Letters* **2017,** 17, (2), 834-841.
8.	Meng, Y.; Zhu, C.; Li, Y.; Yuan, X.; Xiu, F.; Shi, Y.; Xu, Y.; Wang, F. *Opt. Lett.* **2018,** 43, (7), 1503-1506.
9.	Zhou, T.; Zhang, C.; Zhang, H.; Xiu, F.; Yang, Z. *Inorganic Chemistry Frontiers* **2016,** 3, (12), 1637-1643.
10.	Mosca Conte, A.; Pulci, O.; Bechstedt, F. *Scientific Reports* **2017,** 7, 45500.
11.	Gao, J.; Cupolillo, A.; Nappini, S.; Bondino, F.; Edla, R.; Fabio, V.; Sankar, R.; Zhang, Y.-W.; Chiarello, G.; Politano, A. *Advanced Functional Materials* 0, (0), 1900965.
12.	Zhu, C.; Wang, F.; Meng, Y.; Yuan, X.; Xiu, F.; Luo, H.; Wang, Y.; Li, J.; Lv, X.; He, L.; Xu, Y.; Liu, J.; Zhang, C.; Shi, Y.; Zhang, R.; Zhu, S. *Nature Communications* **2017,** 8, 14111.
13.	Jeon, S.; Zhou, B. B.; Gyenis, A.; Feldman, B. E.; Kimchi, I.; Potter, A. C.; Gibson, Q. D.; Cava, R. J.; Vishwanath, A.; Yazdani, A. *Nature Materials* **2014,** 13, 851.
14.	Crassee, I.; Sankar, R.; Lee, W. L.; Akrap, A.; Orlita, M. *Physical Review Materials* **2018,** 2, (12), 120302.
15.	Wang, Z.; Weng, H.; Wu, Q.; Dai, X.; Fang, Z. *Physical Review B* **2013,** 88, (12), 125427.
16.	Li, C.-Z.; Wang, L.-X.; Liu, H.; Wang, J.; Liao, Z.-M.; Yu, D.-P. *Nature Communications* **2015,** 6, 10137.
17.	Zhang, C.; Zhang, E.; Wang, W.; Liu, Y.; Chen, Z.-G.; Lu, S.; Liang, S.; Cao, J.; Yuan, X.; Tang, L.; Li, Q.; Zhou, C.; Gu, T.; Wu, Y.; Zou, J.; Xiu, F. *Nature communications* **2017,** 8, 13741-13741.
18.	Pertsova, A.; Balatsky, A. V. *Physical Review B* **2018,** 97, (7), 075109.
19.	Triola, C.; Pertsova, A.; Markiewicz, R. S.; Balatsky, A. V. *Physical Review B* **2017,** 95, (20), 205410.
20.	Neubauer, D.; Carbotte, J. P.; Nateprov, A. A.; Löhle, A.; Dressel, M.; Pronin, A. V. *Physical Review B* **2016,** 93, (12), 121202.
21.	Carbotte, J. P. *Physical Review B* **2016,** 94, (16), 165111.
22.	Jenkins, G. S.; Lane, C.; Barbiellini, B.; Sushkov, A. B.; Carey, R. L.; Liu, F.; Krizan, J. W.; Kushwaha, S. K.; Gibson, Q.; Chang, T.-R.; Jeng, H.-T.; Lin, H.; Cava, R. J.; Bansil, A.; Drew, H. D. *Physical Review B* **2016,** 94, (8), 085121.
23.	Uykur, E.; Sankar, R.; Schmitz, D.; Kuntscher, C. A. *Physical Review B* **2018,** 97, (19), 195134.
24.	Sharafeev, A.; Gnezdilov, V.; Sankar, R.; Chou, F. C.; Lemmens, P. *Physical Review B* **2017,** 95, (23), 235148.
25.	*APL Materials* **2016,** 4, (12), 126110.
26.	Goyal, M.; Galletti, L.; Salmani-Rezaie, S.; Schumann, T.; Kealhofer, D. A.; Stemmer, S. *APL Materials* **2018,** 6, (2), 026105.
27.	Ali, M. N.; Gibson, Q.; Jeon, S.; Zhou, B. B.; Yazdani, A.; Cava, R. J. *Inorganic Chemistry* **2014,** 53, (8), 4062-4067.
28.	Hosur, P.; Parameswaran, S. A.; Vishwanath, A. *Physical Review Letters* **2012,** 108, (4), 046602.
29.	Wan, X.; Turner, A. M.; Vishwanath, A.; Savrasov, S. Y. *Physical Review B* **2011,** 83, (20), 205101.

30. Liu, Z. K.; Jiang, J.; Zhou, B.; Wang, Z. J.; Zhang, Y.; Weng, H. M.; Prabhakaran, D.; Mo, S. K.; Peng, H.; Dudin, P.; Kim, T.; Hoesch, M.; Fang, Z.; Dai, X.; Shen, Z. X.; Feng, D. L.; Hussain, Z.; Chen, Y. L. *Nature Materials* **2014,** 13, 677.
31. Xu, B.; Dai, Y. M.; Zhao, L. X.; Wang, K.; Yang, R.; Zhang, W.; Liu, J. Y.; Xiao, H.; Chen, G. F.; Taylor, A. J.; Yarotski, D. A.; Prasankumar, R. P.; Qiu, X. G. *Physical Review B* **2016,** 93, (12), 121110.
32. Ashby, P. E. C.; Carbotte, J. P. *Physical Review B* **2014,** 89, (24), 245121.
33. Reddy, H.; Guler, U.; Kildishev, A. V.; Boltasseva, A.; Shalaev, V. M. *Opt. Mater. Express* **2016,** 6, (9), 2776-2802.
34. Ni, G. X.; McLeod, A. S.; Sun, Z.; Wang, L.; Xiong, L.; Post, K. W.; Sunku, S. S.; Jiang, B. Y.; Hone, J.; Dean, C. R.; Fogler, M. M.; Basov, D. N. *Nature* **2018,** 557, (7706), 530-533.
35. Gillen, G. D.; DiRocco, C.; Powers, P.; Guha, S. *Appl. Opt.* **2008,** 47, (2), 164-168.
36. Kresse, G.; Furthmüller, J. *Computational Materials Science* **1996,** 6, (1), 15-50.
37. Kresse, G.; Furthmüller, J. *Physical Review B* **1996,** 54, (16), 11169-11186.
38. Perdew, J. P.; Burke, K.; Ernzerhof, M. *Physical Review Letters* **1996,** 77, (18), 3865-3868.
39. Monkhorst, H. J.; Pack, J. D. *Physical Review B* **1976,** 13, (12), 5188-5192.
40. Togo, A.; Oba, F.; Tanaka, I. *Physical Review B* **2008,** 78, (13), 134106.
41. Omar, M. A., *Elementary Solid State Physics*. 4th ed.; Addison Wesley: 1994.